\input{aipcheck}

\documentclass[
    ,final            
    ,sort&compress
  ]
  {aipproc}

\layoutstyle{6x9}

\usepackage{feynmp}

\newcommand{\mrm}[1]{\ensuremath{\mathrm{#1}}}

\newcommand{\p}{\ensuremath{\mathrm{p}}}

\newcommand{\el}{\ensuremath{{\mathrm{e}}^-}}
\newcommand{\ep}{\ensuremath{{\mathrm{e}}^+}}
\newcommand{\q}{\ensuremath{\mathrm{q}}}
\newcommand{\qbar}{\ensuremath{\bar{\mathrm{q}}}}
\newcommand{\qt}{\ensuremath{\mathrm{t}}}
\newcommand{\qtbar}{\ensuremath{\bar{\mathrm{t}}}}
\newcommand{\g}{\ensuremath{\mathrm{g}}}
\newcommand{\h}{\ensuremath{\mathrm{h}}}
\newcommand{\Z}{\ensuremath{\mathrm{Z}}}
\newcommand{\V}{\ensuremath{\mathrm{V}}}
\newcommand{\W}{\ensuremath{\mathrm{W}}}
\newcommand{\Pythia}{{\small\sf PYTHIA}}
\newcommand{\tsf}[1]{{\small\sf #1}}
\newcommand{\PYTHIA}{\Pythia}
\newcommand{\HERWIG}{{\small\sf HERWIG}}
\newcommand{\ARIADNE}{{\small\sf ARIADNE}}
\newcommand{\SHERPA}{{\small\sf SHERPA}}

\begin{document}

\title{QCD (\&) Event Generators}


\author{Peter Skands}{
  address={Theoretical Physics, Fermilab MS106, Batavia IL-60510-0500,
  USA}
}

\begin{abstract}
Recent developments in QCD phenomenology have spurred on 
several improved approaches to Monte Carlo event
generation, relative to the post--LEP state of the art. In this
brief review, the emphasis is placed on approaches for 1) consistently 
merging fixed--order matrix element calculations with parton showers, 
2) improving the parton
shower algorithms themselves, and 3) improving the description
of the underlying event in hadron collisions.
\end{abstract}

\maketitle


\section{Introduction}
The immediate horizon of accelerator-based high energy experiments is
dominated by HERA, its legacy and final years of running at DESY, 
by the Tevatron, currently in its second run of operations at
Fermilab, and by the Large Hadron Collider, under construction at
CERN. A common denominator for all three machines is the study of high energy
hadronic interactions at unprecedented levels of statistical precision. Thus,
for a wide range of measurements, the limiting factors are ultimately
systematic and theoretical in nature, rather than purely statistical.

Among the most important challenges is naturally that,
while perturbative QCD describes the interactions of
quarks and gluons, experiments observe 
hadrons. In addition, many collider observables involve an interplay
between widely separated energy scales, logarithms of which 
may appear and impact the validity of predictions even 
at the perturbative level. As a result, the field of QCD phenomenology is
experiencing a rapid pace of development, a significant portion of 
which can be traced to either of two sources that I will focus on here. 

Firstly, as the Centre-of-Mass energy increases, 
the phase space for radiation also becomes larger; 
high--$\hat{s}$ final states are likely to be
accompanied by high jet multiplicities. To accurately predict 
observables in such processes, 
lowest order scattering matrix elements are not sufficient.
Rather, more sophisticated approaches are
called for, which combine the rate of hard wide--angle 
jets predicted by fixed--order $2 \to n$ matrix elements 
with a resummation of multiple soft emissions in a consistent way, avoiding
both double counting and ``dead regions'' over all of phase space. 

Secondly, hadron collisions involve new intrinsic challenges relative to 
$\mathrm{e}\mathrm{e}$ (and to some extent also
$\mathrm{e}\mathrm{p}$) scatterings, since
both of the initial states are here composite and strongly
interacting. However, on the
theoretical side, the description of beam remnants and underlying
events has gone through a long period of relative
hibernation, essentially during the LEP era, with few new ideas
emerging over the last $\sim20$ years. Recently, however, 
interest has been rekindled, largely in response to 
increased interest from the Tevatron and LHC collaborations. 

Staying along the same lines, it should also here be emphasized that,
when using  
LEP, HERA, and Tevatron results to make predictions for the LHC, an
extrapolation is performed over orders of magnitude in $Q^2$ and $x$, 
at the same time as  approximations 
that are ``safe''  
at lower energies may be stretched into regions where large corrections are 
to be expected. As such, a non-trivial and many-faceted issue is how to
treat the associated systematic and theoretical uncertainties. 
Though I will not touch directly on this topic below, some recent 
progress that partly addresses it 
is the emergence of parton distributions with intrinsic
errors, reported on elsewhere in these proceedings
\cite{dis_pumplin,dis_stump,dis_thorne}.  

\section{Hard \& Soft -- ME/PS Matching}
The evaluation of tree--level 
transition amplitudes, involving less than, say, 5--6
partons in the final state, is a procedure which by now has been
largely automated. Matrix Element Generators like \tsf{CompHEP}\
\cite{Pukhov:1999gg}, \tsf{Grace}\ \cite{Yuasa:1999rg,Tsuno:2002ce}, 
\tsf{HELAS} \cite{Murayama:1992gi}, \tsf{MadGraph}\
\cite{Stelzer:1994ta,Maltoni:2002qb}, \tsf{O'Mega}\
\cite{Moretti:2001zz},  
and \tsf{AMEGIC}++ \cite{Krauss:2001iv}
provide fast and reliable means of
obtaining (more or less) tractable analytic expressions for a broad
range of matrix elements, both in the Standard Model and beyond. 
Combining these with efficient numerical phase space
integration algorithms such as \tsf{BASES/SPRING}\
\cite{Kawabata:1995th} and others 
\cite{Ilyin:1996gy}, it is possible to further automate the phase
space weighting, and hence to generate events corresponding to the
chosen amplitude at matrix--element level. 
A number of more dedicated matrix element evaluation codes also exist, 
with processes hard-coded one by one, most notably 
\tsf{AlpGen}\ \cite{Mangano:2002ea}, but also \tsf{MCFM}\ 
\cite{Campbell:2000bg}
 and others 
 \cite{Beenakker:1996ed,Berends:1989ie,Berends:1990ax,Giele:1993dj,Kersevan:2002dd,Nagy:1998bb}. 

Common to all these approaches (most at tree level and a few at one
loop) is that they represent fixed--order
expansions in the electromagnetic, weak, and in particular strong
coupling constants. As such, the virtue of these calculations is, briefly
stated, that they include the entire helicity and interference
structure of the amplitude (as well as virtual corrections to it, to
the extent that loops are included), up to the given order calculated. 
Furthermore, asymptotic freedom implies that the stability of this
expansion should improve with energy, due to the gradual vanishing of
the strong coupling at large energies.
Admittedly, the complexity rapidly grows with the number of
particles involved, and so as already mentioned, these approaches are
presently limited to fairly inclusive observables, 
where the number of resolved 
final state particles does not exceed a handful or
so.

On the other hand, from LEP we know that multiple soft gluon emissions are
important in building up the full event structure. Mathematically,
these corrections correspond to 
logarithmic enhancements of the amplitude, of the form
\begin{equation}
\alpha_s^N
\log^{2N}(Q_{\mathrm{hard}}/Q_{\mathrm{soft}}) + ...
\end{equation}
where $Q_\mathrm{soft}/Q_{\mathrm{hard}}$ is a measure of the softness of the
gluon(s) relative to the hard scale(s) in the problem.
Moreover, when going to higher energies, the phase space for such emissions
increases. Thus,
while fixed--order calculations should be able to predict reliably the
rate of a few well--separated jets (and other observables at the same level
of inclusiveness), it is necessary go beyond the fixed--order
approximation to obtain a picture of the full event structure.

To improve the logarithmic accuracy, two dominant approaches exist:
resummation calculations, and parton showers. Both are approximations
to perturbation theory which work at infinite order in the coupling
constant, and which are exact in certain limits. 

The former approach, resummation, 
allows to include not only the terms
shown explicitly above (double logs $\sim$ LLA), but also less
singular logarithms in  a systematic way. However, 
the formalism can still only be applied to relatively
inclusive quantities, and a separate calculation must be performed for
each observable, though interesting work has recently been 
carried out on automating calculations of this type \cite{Banfi:2004yd}. 

In this talk, I will concentrate on the parton shower approach. 
While this description is formally correct only 
to leading logarithmic accuracy, it has the virtue that a fully
exclusive description of the final state is obtained, which can be
easily matched onto hadronisation descriptions, and from which in
principle any observable can be constructed. Moreover,
it is possible (and indeed necessary, as e.g.\ in the case of momentum
conservation) to include at least a subset of higher--order effects, 
such as angular ordering of emissions, optimizing the scale choice in 
$\alpha_s$ with respect to higher order kernels \cite{Amati:1980ch}, 
and choosing azimuthal angles in the branchings non--isotropically
\cite{Webber:1986mc}. In practice,
such refinements have been introduced in all of the standard shower
Monte Carlos, including in particular the \ARIADNE\ dipole shower
\cite{Gustafson:1987rq,Lonnblad:1992tz,Lonnblad:1995ex}, the \HERWIG\ 
\cite{Marchesini:1983bm,Seymour:1994df,Corcella:2000bw} and \HERWIG++ 
\cite{Gieseke:2003rz} showers,
 and both the \PYTHIA\ virtuality--ordered 
 \cite{Bengtsson:1986et,Norrbin:2000uu,Sjostrand:2000wi} and
 transverse-momentum--ordered 
 \cite{Sjostrand:2004ef} showers  
 (the \SHERPA\ generator \cite{Gleisberg:2003xi} basically 
 uses a variant of the \PYTHIA\ virtuality--ordered algorithm). It 
would therefore be grossly misleading to equate 
leading log analytical calculations, where no such refinements are
 included, with leading log parton showers. 

The parton shower approximation starts from the observation that the
collinear limit of QCD (and QED, for that matter) is universal. Thus, a process
like $\mathrm{e}^+\mathrm{e}^-\to\mathrm{q}\bar{\mathrm{q}}$ can be
corrected to  the process 
$\mathrm{e}^+\mathrm{e}^-\to\mathrm{q}\bar{\mathrm{q}}\mathrm{g}$ 
using universal expressions for the $\mathrm{q}\to\mathrm{q}\mathrm{g}$
splitting probability. Due to the universality, the same expressions
may then be applied again to describe the radiation of further gluons,
as well as gluon splittings into quarks and so forth.
Since the integrated probability at each step is nominally
infinite, an ordering is introduced, whereby the emissions are
generated sequentially according to some resolution criterion, 
like angle, virtuality, or transverse momentum. 
A lower cutoff on the resolution variable may
then be naturally introduced, that regulates the infrared divergences,
and at which scale a hadronisation description is supposed to take over. 

Thus, the virtues are that final states with an arbitrary number of partons may
be built up, with a transition to hadronisation
descriptions built in from the start. The down side is that the
approximation is only exact in the collinear limit. For hard and/or
wide--angle emissions, different parton showers can give widely
different answers, reflecting the approximate nature of the approach
in those regions. However, as mentioned above these are precisely 
the regions where the fixed--order calculations are at their best,
hence it has been a long--standing wish to join consistently the state
of the art of both worlds.

\subsection{ME/PS Merging}
The simplest (and oldest) approaches to join matrix elements
and parton showers I will here refer to as
matrix--element/parton--shower (ME/PS) ``merging'', to be
contrasted with ``matching'' below. 
Essentially, merging
improves the parton shower off a hard system, call it $X$, by re-weighting the
position of the hardest jet in phase space to reproduce the matrix element
distribution for $X$+jet. An overview of hadron collider 
processes for which such
corrections have been implemented in the \HERWIG\ and \PYTHIA\ models is
given in Tab.~\ref{tab:merging}. 
\begin{table}
\begin{tabular}{lcccccc}
\hline
\tablehead{1}{r}{t}{}  & 
  \tablehead{1}{r}{t}{$\mathrm{pp}\to \mathrm{h}^0$ }
& \tablehead{1}{r}{t}{$\mathrm{pp}\to \mathrm{V}$ ($=\gamma/\mathrm{Z}/\mathrm{W}$)}
& \tablehead{1}{r}{t}{DIS} 
& \tablehead{1}{r}{t}{top decay } 
& \tablehead{1}{r}{t}{SM decays } 
& \tablehead{1}{r}{t}{SUSY decays\tablenote{corrections approximate
for 2--body RPV modes, absent for RPV 3--body modes
\cite{Sjostrand:2002ip}.}} \\ 
\hline
\HERWIG & $\surd$ & $\surd$ & $\surd$ & $\surd$ & - & - \\
\PYTHIA\tablenote{{PYTHIA}: applies to both the $Q^2$ and $p_\perp^2$
ordered shower algorithms in PYTHIA 6.3.}   
& $\surd$& $\surd$ & $\surd$ & $\surd$ & $\surd$& $\surd$ \\
\hline
\end{tabular}
\caption{{\bf ME/PS Merging:}
List of processes, $X$, for which $X$+jet merging
has been implemented in the \HERWIG\ and \PYTHIA\ Monte Carlos
(\ep\el\ processes omitted).
\label{tab:merging}}
\end{table}

Technically, the way these corrections are implemented can be quite
different, depending on the showering algorithm. In \HERWIG, the 
showering algorithm has a ``dead zone'' 
in the hard wide--angle region, where no radiation at all is
produced. In order to match to the matrix element which does produce
jets there, two classes of events are effectively merged, 
e.g.\ $X$ and $X$+jet, with the latter chosen inside the dead region of
the former \cite{Seymour:1994df}. 
The detailed procedure is somewhat complicated and has
only been worked out for a few cases, most recently for Higgs
production \cite{Corcella:2004fu}.

In \PYTHIA, the problem is rather the opposite. Too much radiation is
generally produced in the hard wide--angle region, as compared to the
matrix element answer. It is thereby straightforward to introduce a
re-weighting, vetoing some of the extra emissions, to arrive back down
at the matrix element rate \cite{Norrbin:2000uu,Miu:1998ju}. 
Note that these corrections 
are applied for  both the $Q^2$-- and $p_{\perp}^2$--ordered
shower algorithms in \PYTHIA\ 6.3.

So far, so good. However, for both the \HERWIG\ and \PYTHIA\ style
merging, the procedure rapidly becomes more involved when
attempting to generalize the methods to more complicated final
states (see e.g.\ \cite{Andre:1997vh,Mrenna:1999mq}). 
Moreover, recent developments along related 
lines have resulted in a range of more generic approaches which are
now being more actively pursued, as will be discussed below.

\subsection{ME/PS Matching at Leading Order}
The problem of consistently adding parton showers to 
a set of leading--order matrix elements for $X$, $X$ + jet, $X$ + 2
jets, etc, to obtain an inclusive sample of $X$ production, matched to
all available hard radiation matrix elements, has recently been studied in
detail by a number of authors, in particular by Mangano (\tsf{MLM})
\cite{Mangano}, by Catani, Krauss,
Kuhn, and Webber (\tsf{CKKW}) \cite{Catani:2001cc,Krauss:2002up}, by
L\"{o}nnblad  
\cite{Lonnblad:2001iq}, and most recently 
by Mrenna and Richardson \cite{Mrenna:2003if}. 
(See also talk by Frixione \cite{dis_frixione}.) 

All these approaches essentially allow a consistent adding together of
events generated with different jet multiplicities at the
matrix--element level (e.g.\ $\W$, $\W$+jet, $\W$+2jets, ...), by 
re-weighting them and showering them in way so that
double counting and empty regions are avoided over all of phase space. 

The approach proposed by \tsf{CKKW} \cite{Catani:2001cc} is, briefly stated,
to first select the jet multiplicity, $n$, at the matrix
element level according to a known probability, 
\begin{equation}
P(n) = \frac{\sigma_n}{\sigma_0+\sigma_1+\sigma_2+\sigma_3 + ...}, ~~~~~~~ ; ~~~~ \sigma_i \equiv \sigma_i(Q_{\mrm{cut}}),
\end{equation}
with $Q_{\mrm{cut}}$ a cutoff scale (in principle arbitrary) 
that regulates the infrared divergencies 
of the matrix elements. According to the chosen matrix element,  
a set of explicit four-momenta $p_i$ are then generated, to which 
a jet clustering algorithm is applied. A series of 'branchings' is
thereby reconstructed, which can be interpreted as a parton shower
history. The event as a whole is then re-weighted according to the Sudakov form
factors (see below) and $\alpha_s$ values associated with the reconstructed 
intermediate scales. 
A parton shower can then be applied as the final step, with
emissions above the cut scale vetoed for all except the highest jet
multiplicity matrix elements.

\emph{Why} this works is more technical: recall that the parton shower
is formulated in terms of the no--emission
probability between two scales, the Sudakov form factor, which in 
all simplicity is the (singular part of the) 
probability for an $n$--jet configuration to remain an $n$--jet
one as a function of the resolution scale. By 
re-weighting the matrix elements with Sudakov factors for each leg, 
the leading divergencies which would lead to double--counting between e.g.\
$\sigma_1$ and $\sigma_2$ are cancelled. Roughly speaking, the Sudakov
re-weighting  
takes into account that for every $2$-jet configuration you gain, you
must lose one $1$-jet one. It was shown by \cite{Catani:2001cc} that this
procedure makes the sum stable at least to next-to-leading logarithmic (NLL)
accuracy. Note that this stabilisation is to some extent 
equivalent to the cancellation of
real divergencies by virtual ones in a full NLO calculation, 
with the Sudakovs here
 playing the role of virtual corrections, that have the same structure
(but opposite signs) as the tree--level divergencies. 

To further explain what the Sudakovs are doing, note that the parton shower 
is correct in the limit of strongly ordered emissions, i.e.\ 
in the limit that each successive scale is much smaller than
the preceding one. In this case, the emission probability is
large (i.e.\ the Sudakov = \emph{no}--emission probability, is small), 
and the Sudakov re-weighting becomes quite important. At the other
extreme are emissions which happen at similar scales; here,
the Sudakovs are very close to one (the probability to go from one
scale to another without emission becomes unity in the limit that the
two scales coincide), and hence the
matrix element results remain essentially unaltered here, as desired. 

This style of matching has since
been implemented in the \SHERPA\ event generator for several
processes \cite{Gleisberg:2003xi}. 
It was then noticed by L\"onnblad \cite{Lonnblad:2001iq} 
that a better matching could be
obtained by replacing the analytical Sudakovs used in the original
\tsf{CKKW}\ prescription by Sudakovs numerically
generated by running the actual parton shower (so--called
`pseudo--showers'). This is the approach 
implemented \cite{Lavesson:2005xu} in the \ARIADNE\
generator \cite{Lonnblad:1992tz}. 
Mrenna and Richardson subsequently made further refinements
\cite{Mrenna:2003if}, applying the methodology also to hadronic
collisions, using
\tsf{MadGraph}\ and the \HERWIG\ and \PYTHIA\ generators, part of
which work is stored in the \tsf{PATRIOT}\ event database \cite{patriot}. 

Mangano's prescription (``\tsf{MLM}\ matching'')  \cite{Mangano}
is similar but somewhat simpler in spirit than \tsf{CKKW}. In
particular, it is based on clustering of 
events after showering and thus has a much simpler interface between
the matrix element and parton shower generators. 

\begin{table}
\begin{tabular}{lccccc}
\hline
\tablehead{1}{r}{t}{}  & 
  \tablehead{1}{r}{t}{$\mathrm{\ep\el}\to \q\qbar$}
& \tablehead{1}{r}{t}{$\mathrm{pp}\to \mathrm{V}$
($=\gamma/\mathrm{Z}/\mathrm{W}$)} 
& \tablehead{1}{r}{t}{$\mathrm{pp}\to \mathrm{VV}$} 
& \tablehead{1}{r}{t}{DIS} \\
\hline
\ARIADNE\ & $\surd$ & $\surd$\tablenote{$\V=\W$ only.} &  - & $\surd$\\
\SHERPA & $\surd$ & $\surd$ & $\surd$ & - & \\
\tsf{PATRIOT} & - & $\surd$ & - & - & \\
\hline
\end{tabular}
\caption{{\bf LO ME/PS Matching:}
List of processes, $X$, for which LO ME/PS matched generators/samples
are available.
\label{tab:lomatching}}
\end{table}
Tab.~\ref{tab:lomatching} gives an overview of processes for which
Leading Order matching is currently implemented/available.

\subsection{ME/PS Matching at NLO}
Above, the aim was essentially to describe real QCD radiation as
accurately as possible, over all regions 
of phase space. This was accomplished by consistently matching on
parton shower descriptions of 
soft radiation to a set of tree--level matrix elements
describing as many hard emissions as one cares to calculate, which in
practice currently means up to 3--4 extra jets.

However, since virtual corrections are not included, the
normalization of the distributions is still only correct to leading
order. Quite recently, the problem of matching parton showers to the
full NLO theory, i.e.\ one--leg \emph{and} one--loop corrections to the
lowest order, has been addressed by several
groups (see also talk by Frixione \cite{dis_frixione}). 
Early approaches include the use of phase space slicing to
separate resolvable and unresolvable regions
\cite{Dobbs:2001dq,Potter:2001ej}, which, despite a number of
initial successes, suffers from the drawback of not reproducing the
perturbative expansion correctly. Important studies 
have also been carried out by the group
\cite{Collins:2000qd,Collins:2000gd,Collins:2004vq}, though so far
practical applications have been limited.

Presently, the most mature NLO matching approach is the one put forth
by Frixione, Nason, and Webber
\cite{Frixione:2002ik,Frixione:2003ei}, which has been implemented in
the program \tsf{MC@NLO} (essentially a superstructure built
onto the \HERWIG\ Monte Carlo). Another promising approach 
suggested by Kr\"amer and Soper \cite{Kramer:2003jk,Soper:2003ya} 
has so far only
been applied to $\ep\el$ observables, though work is in progress to
generalise it. A more complete list of processes is given in
Tab.~\ref{tab:nlomatching}.  
\begin{table}
\begin{tabular}{lcccccc}
\hline
\tablehead{1}{r}{t}{}  & 
  \tablehead{1}{r}{t}{$\mathrm{\ep\el}\to \q\qbar$
}
& \tablehead{1}{r}{t}{$\mathrm{pp}\to \mathrm{h}^0$}
& \tablehead{1}{r}{t}{$\mathrm{pp}\to \mathrm{V}$
($=\gamma/\mathrm{Z}/\mathrm{W}$)} 
& \tablehead{1}{r}{t}{$\mathrm{pp}\to \mathrm{VV}$} 
& \tablehead{1}{r}{t}{$\mathrm{pp}\to \mathrm{QQ}$} 
& \tablehead{1}{r}{t}{single top} \\\hline
\tsf{MC@NLO} & -\tablenote{Note that, due to the simple structure of
\ep\el, NLO matching can here effectively be obtained 
using the existing HERWIG and PYTHIA merging
methods, and re-weighting by the loop factor (1+$\alpha_s/\pi$).}
& $\surd$ & $\surd$ & $\surd$ &  $\surd$\tablenote{$Q$:
heavy quark.} & in progress\\
\hline
\end{tabular}
\caption{{\bf NLO ME/PS Matching:}
List of processes, $X$, for which unweighted NLO ME/PS matched event
samples can
be generated with the \tsf{MC@NLO} generator (weights = $\pm 1$).
\label{tab:nlomatching}}
\end{table}

Naturally, the big boon is that one 
automatically obtains cross sections normalised to NLO precision. 
A vice, as compared to the leading order matchings, 
has so far been that the NLO matrix elements only include 
the first (real) hard emission; subsequent emissions, even when hard,
must still be generated by the parton shower, though recent work has
been carried out on including also \tsf{CKKW}-style matching for
higher jet multiplicities \cite{Nagy:2005aa}.  

A related issue is that, since \tsf{MC@NLO}\ is hard-wired to \HERWIG, 
it is presently not possible to vary the parton shower model. 
Consider for instance the peak position of the Drell--Yan
and $\h^0$  $p_\perp$ spectra. This is where the bulk of the cross section
sits, and this is, roughly speaking, the region that gets the most 
enhancement by the loop corrections. However, this region is also
highly sensitive to multiple soft emissions, which are resummed by the
parton shower. A much awaited future development is thus the
creation of tools that allow a more generic interfacing with different
shower models. 

In a similar vein, note that, while the overall normalisation
is improved by going from LO to NLO matching, the same is not
necessarily true for the shapes of distributions, as follows.
Consider again the Drell--Yan $p_\perp$ spectrum in hadron collisions. 
As mentioned above, it is dominated by $\Z$ + multiple 
soft emissions in the peak, while $\Z$+jet dominates in the tail. 
It would now
be possible experimentally to define a semi-inclusive $\Z$+jet 
\emph{fraction} as a function of $p_{\perp\Z}$. At NLO, the
prediction for $Z$+jet (and hence for this subsample) is 
effectively only correct to leading order, 
since virtual corrections to $Z$+jet are not included. Since 
the $Z$+jet fraction tends towards unity at large $p_{\perp\Z}$ values, 
the overall shape should probably not 
be regarded as being more precisely determined here than in the leading
order matching schemes discussed above. 
The extreme example would  be an observable
sensitive to multiple hard emissions, where a leading order matching
including several hard jet radiations would clearly be superior to the
present NLO matching schemes, 
where the second jet has to be radiated by the parton shower. 

\section{New Parton Shower Algorithms}
Both \tsf{HERWIG}++ \cite{Gieseke:2003hm} 
and \tsf{PYTHIA} 6.3  \cite{Sjostrand:2003wg} 
contain new parton
shower models. In the \tsf{HERWIG}++ case, refinements have been made
\cite{Gieseke:2003rz} 
on the already existing \HERWIG\ model, 
while a completely new shower
model  \cite{Sjostrand:2004ef} has been implemented in 
\tsf{PYTHIA} 6.3. The recently completed program \tsf{APACIC}++ 
\cite{Krauss:2005re} also contains a parton shower model, which
essentially is an adaption of the old \tsf{PYTHIA} shower model with
the specific implementation of \tsf{CKKW} style matching in mind.

The \HERWIG\ shower is based on a strictly angular--ordered sequence
of emissions \cite{Marchesini:1983bm}. This correctly accounts for
coherence effects in the emission of soft gluons, but has the
disadvantage that it leaves a  `dead zone' in the hard 3--jet region,
which has to be filled in separately, as discussed above. The shower
evolution is stopped once a fixed low scale is reached, at which time a
transition is made to a non--perturbative hadronisation description,
the \HERWIG\ one being based on the cluster model 
\cite{Webber:1983if}.

The \HERWIG++ algorithm \cite{Gieseke:2003rz} 
starts from the same basic principles and thus
inherits most features from the old shower, 
but the definition of the evolution variable 
has been slightly changed, to allow a more correct
treatment of heavy quark radiation as well as a more consistent
behaviour in the soft--gluon limit. Finally, the treatment of cascade
decays of resonances interspersed with showers should be improved by a
more consistent implementation of multi--scale showers. 
Recent studies exploring this algorithm can be found in 
\cite{Gieseke:2003rz,Gieseke:2004af,Gieseke:2004tc}. At present,
\HERWIG++ is still mainly a tool for \ep\el\ collisions, though work is 
underway to extend it to hadron collisions. Another 
item on the agenda is the inclusion of a \tsf{CKKW}-style ME/PS
matching scheme. 

More radical changes have occurred in \PYTHIA\ 6.3, where a new
transverse--momentum--ordered shower based on a recently proposed 
hybrid between the dipole and parton shower formalisms
\cite{Sjostrand:2004ef} has been implemented (in addition to the old
virtuality--ordered shower, which is carried over from previous
versions). The choice of $p_\perp$ as evolution variable was made for
several reasons. Firstly, it has the dual property of simultaneously
being a good measure of hardness while still leading to a natural angular
ordering of emissions \cite{Gustafson:1986db}. Secondly, it is Lorentz
invariant under longitudinal boosts, which is not the case e.g.\ for the
\HERWIG\ angular--ordered evolution variable. Thirdly, while both
\PYTHIA\ and \HERWIG\ can be tuned to give good descriptions of the
LEP data, the \tsf{ARIADNE}\ $p_\perp$--ordered dipole shower 
tends to do even better. Note, however, that the $p_\perp$ measure
proposed in \cite{Sjostrand:2004ef} is different from the one used in 
the \tsf{ARIADNE}\ evolution. Fourthly, the underlying--event model in
\tsf{PYTHIA} is based on a $p_\perp$--ordered sequence of multiple
perturbative interactions, hence a shower also ordered in $p_\perp$
allows for a more unified treatment of the perturbative activity as a
whole, as will be discussed below. 

\begin{figure}[t]
\begin{fmffile}{fsrbranch}
\begin{fmfgraph*}(200,120)
\fmfleft{z1}
\fmfright{b,m2,mt,m1,m0,t}
\fmf{boson,tension=5}{z1,z2}
\fmf{fermion,tension=2,label=$\q^*$,label.side=left}{z2,f1}
\fmf{fermion}{f1,t}
\fmfv{label=$\q$,l.ang=0,l.dist=4}{t}
\fmf{phantom}{m1,f1}
\fmf{fermion,tension=2,label=$\qbar^*$,label.side=left}{f2,z2}
\fmf{fermion}{b,f2}
\fmfv{label=$\qbar$,l.ang=0,l.dist=4}{b}
\fmf{phantom}{m2,f2}
\fmfv{label={$P_{\q\to\q\g}(z)$},l.ang=0,l.dist=3}{f1}
\fmfv{label={$P_{\qbar\to\qbar\g}(z)$},l.ang=0,l.dist=3}{f2}
\fmfv{l.ang=-100,l.dist=29,label=\parbox{3cm}{$p_{\perp\mrm{evol}}^2 \equiv$\\$
      z(1-z)(m_*^2-m_0^2)$}}{z2}
\fmfv{label=$\hat{s}$}{z1}
\fmfv{d.sh=circ,d.siz=5}{f1}
\fmfv{d.sh=circ,d.siz=5}{f2}
\fmffreeze
\fmf{plain,left=0.2}{f1,f2}
\fmf{plain}{f1,g,f2}
\fmf{plain,right=0.2}{f1,f2}
\fmffreeze
\fmf{gluon}{g,m1}
\fmf{plain,left=0.2,foreground=(0.83,,0.83,,0.83)}{b,m1}
\fmf{plain,foreground=(0.83,,0.83,,0.83)}{b,m1}
\fmf{plain,right=0.2,foreground=(0.83,,0.83,,0.83)}{b,m1}
\fmf{plain,left=0.2,fore=(0.83,,0.83,,0.83)}{t,m1}
\fmf{plain,fore=(0.83,,0.83,,0.83)}{t,m1}
\fmf{plain,right=0.2,fore=(0.83,,0.83,,0.83)}{t,m1}
\end{fmfgraph*}
\end{fmffile}
\caption{Final--state branching in the
$p_\perp$--ordered shower. The evolution is performed on each side of
the dipole separately, in the variable
$p_{\perp\mrm{evol}}^2$, here expressed in terms of the parton
virtuality, $m_*$, its rest mass, $m_0$ (for massive partons), and the energy
sharing fraction $z$ appearing in the splitting kernels
$P(z)$. Kinematics are constructed in the 
dipole CM frame, conserving energy and momentum inside the system.
\label{fig:ptevolution}
}
\end{figure}
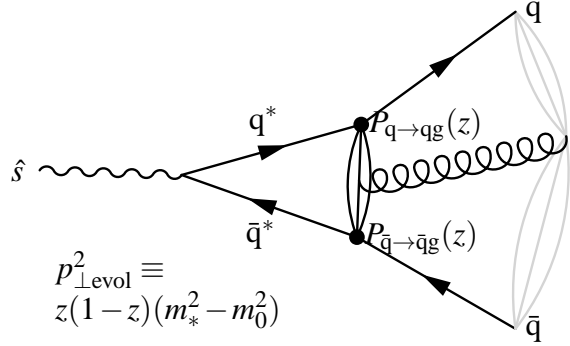

The dipole/parton shower hybrid approach implies an evolution of 
individual parton lines, but with recoils occurring inside dipoles, 
as illustrated in Fig.~\ref{fig:ptevolution}. 
For the final--state shower, 
these dipoles are normally defined by the
respective colour partners, with some exceptions involving decays of heavy
resonances. For the initial--state shower, where a backwards evolution is
performed from the hard scattering down, only a single dipole is
relevant, spanned between the two incoming partons.

\begin{figure}
\includegraphics*[scale=1.0]{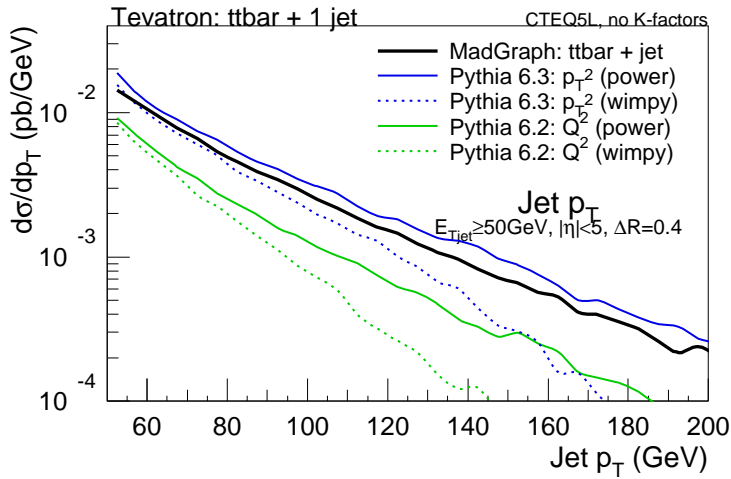}
\caption{The Tevatron $\qt\qtbar$+jet rate as a function of jet
$p_\perp$. Results are shown for
MadGraph, and for hard (`power') and soft (`wimpy') 
variants of both the $Q^2$--
and $p_\perp^2$--ordered PYTHIA showers, see text.
\label{fig:ttbar}
}
\end{figure}

Studies carried out so far indicate that the new shower leads to an
improved description e.g.\ of event shapes at LEP \cite{Rudolph} and of 
Drell--Yan production at the Tevatron
\cite{Sjostrand:2004ef,Huston:2004yp}. Fig.~\ref{fig:ttbar} gives a
preliminary comparison of the $\qt\qtbar$+jet rate at the Tevatron, 
as a function of jet $p_\perp$ \cite{plehnprep}. Results are shown for  
\tsf{MadGraph}\ (thick black line), for two variants  of the old
$Q^2$--ordered shower (green lines, solid and dotted),
and for two variants of the new $p_\perp^2$--ordered shower (blue
lines, solid and dotted). 
The 'power' and 'wimpy' versions of each
shower represent a reasonable range of variation of the maximum scale
for shower emissions, with the former populating all available phase
space, and the latter bounded by the scale of the hard interaction.
 Though it is still too early to draw more general conclusions, 
it is encouraging that both the shape and normalisation of
the distribution appear to be improved, especially in the 'power
shower' variant of the $p_\perp$--ordered algorithm.

\section{The Underlying Event}
The underlying event may be defined somewhat loosely 
as the activity in a (single)
hadron--hadron collision which does not originate directly from the
hard scattering that triggered the event. Currently, this component
is not well understood from first principles. What is known, on the
other hand, is that it produces omnipresent systematic
as well as random fluctuations in activity, which impact isolation
criteria, jet energy scales, etc., which  can be of significant
magnitude for experimental analyses. 
 
During the last few years it has become increasingly clear that the
underlying event not only contains soft activity but also has a 
semi--hard, 'lumpy' component \cite{Field}. To explain this, the 
concept of multiple perturbative interactions \cite{Sjostrand:1987su} 
appears increasingly attractive. Several recent implementations are
built on ideas incorporating multiple interactions in one form or
another, including the underlying--event framework in the \HERWIG\
add--on \tsf{JIMMY} \cite{Butterworth:1996zw}, 
the new interleaved model \cite{Sjostrand:2004pf,Sjostrand:2004ef} 
in \tsf{PYTHIA} 6.3, 
and the underlying--event model being developed for 
\tsf{SHERPA} \cite{uesherpa}, this latter  
being quite similar to the old \tsf{PYTHIA} one \cite{Sjostrand:1987su}.

The most recent development has been the interleaved model of
\cite{Sjostrand:2004pf,Sjostrand:2004ef}. In this picture, the
evolution of the initial state radiation and the generation of
multiple scatterings are no longer independent. Instead, a successive
`fine-graining' of all the perturbative activity is performed, with
radiations interspersed (or `interleaved') with interactions. This
allows correlations to be introduced between all the perturbative
activity at successively smaller scales. For instance, the third
interaction will know about the presence of a gluon having been
radiated off a valence quark in the
first, etcetera. This introduces for the first time non--trivial
correlations in flavour and $x$, while all known sum rules are still
respected (e.g.\ momentum conservation and quark counting rules).
In addition, the model allows new possibilities for 
impact parameter dependence, and contains a more refined treatment of
beam remnants, based on an extension of the Lund string model to
`baryonic' colour topologies \cite{Sjostrand:2002ip}. 

With the development of more sophisticated physics models, 
the hope is that it will 
be possible to ask a range of more 
meaningful physics questions, among which especially the topic of colour
correlations and colour reconnections (in view of a possible
difference between the vacua left by \ep\el\ and hadronic initial
states) is presently the most actively investigated. Finally, the
energy dependence of the underlying activity is currently very 
poorly known; further
studies aiming to pin down the scaling behaviour, for
instance by including the available RHIC \p\p\ data, 
would be of great interest.

\section{Conclusion}
The quest for the cause of electroweak symmetry breaking is 
(hopefully) nearing
its end. Should the Tevatron not discover it within a few years, the
experimental programme at the LHC promises a comprehensive exploration
of the TeV scale, with the emphasis on precisely this question, as
well as on more general searches for signals of physics beyond the
Standard Model. In addition, 
the high energies and hadronic environment of both the
Tevatron and in particular the LHC will challenge 
our understanding of QCD, especially our
ability to solve it for large 
numbers of partons, both real and virtual, and our control of
phenomena near the borders of its perturbative domain. 

It is encouraging, then, to see a truly impressive effort 
mounted in recent years to address several of the most important 
issues, among which I have here discussed 
more precise and realistic simulation of events at higher perturbative
orders, improved parton shower models, and some progress towards a
better understanding of the underlying event. 

On a general note, the emergence of many sophisticated but 
specialised tools entails an increased need for efficient 
cross--communication between the programs. 
For instance, while the traditional Monte Carlo generators have been 
largely self-contained in the past, the time-consuming process of 
hard--coding matrix elements for each process by hand
can now be left to automatic programs optimised specifically for this task.
One should therefore expect an increased reliance 
on external interfaces, such as provided by 
the Les Houches Accords \cite{Boos:2001cv,Skands:2003cj}, 
in the future. 

Finally, though much of the effort reported here is still centred around a
relatively few groups, for which simple manpower restrictions often
represent a non--trivial problem, 
the many new and exciting
developments do seem to have led to an increased communication, bringing 
people together from different fields. Hopefully, with continued 
nourishment, this is a trend that will continue and grow in the future.






\bibliographystyle{aipproc}   

\bibliography{sample}

\hyphenation{Post-Script Sprin-ger}
\begin{thebibliography}{76}
\expandafter\ifx\csname natexlab\endcsname\relax\def\natexlab#1{#1}\fi
\providecommand{\enquote}[1]{``#1''}
\expandafter\ifx\csname url\endcsname\relax
  \def\url#1{\texttt{#1}}\fi
\expandafter\ifx\csname urlprefix\endcsname\relax\def\urlprefix{URL }\fi
\providecommand{\eprint}[2][]{\url{#2}}

\bibitem[Pumplin(2005)]{dis_pumplin}
J.~Pumplin, \enquote{Parton Distributions,} in \emph{these proceedings}, 2005,
  transparencies available from www.hep.wisc.edu/dis05/.

\bibitem[Stump(2005)]{dis_stump}
D.~Stump, \enquote{Stability and uncertainty of parton distribution functions,}
  in \emph{these proceedings}, 2005, transparencies available from
  www.hep.wisc.edu/dis05/.

\bibitem[Thorne(2005)]{dis_thorne}
R.~Thorne, \enquote{MRST PDFs and impact on LHC physics,} in \emph{these
  proceedings}, 2005, transparencies available from www.hep.wisc.edu/dis05/.

\bibitem[Pukhov et~al.(1999)]{Pukhov:1999gg}
A.~Pukhov, et~al. (1999), \eprint{hep-ph/9908288}.

\bibitem[Yuasa et~al.(2000)]{Yuasa:1999rg}
F.~Yuasa, et~al., \emph{Prog. Theor. Phys. Suppl.}, \textbf{138}, 18--23
  (2000), \eprint{hep-ph/0007053}.

\bibitem[Tsuno et~al.(2003)]{Tsuno:2002ce}
S.~Tsuno, et~al., \emph{Comput. Phys. Commun.}, \textbf{151}, 216--240 (2003),
  \eprint{hep-ph/0204222}.

\bibitem[Murayama et~al.(1991)]{Murayama:1992gi}
H.~Murayama, I.~Watanabe, and K.~Hagiwara (1991), kEK-91-11.

\bibitem[Stelzer and Long(1994)]{Stelzer:1994ta}
T.~Stelzer, and W.~F. Long, \emph{Comput. Phys. Commun.}, \textbf{81}, 357--371
  (1994), \eprint{hep-ph/9401258}.

\bibitem[Maltoni and Stelzer(2003)]{Maltoni:2002qb}
F.~Maltoni, and T.~Stelzer, \emph{JHEP}, \textbf{02}, 027 (2003),
  \eprint{hep-ph/0208156}.

\bibitem[Moretti et~al.(2001)]{Moretti:2001zz}
M.~Moretti, T.~Ohl, and J.~Reuter (2001), \eprint{hep-ph/0102195}.

\bibitem[Krauss et~al.(2002)]{Krauss:2001iv}
F.~Krauss, R.~Kuhn, and G.~Soff, \emph{JHEP}, \textbf{02}, 044 (2002),
  \eprint{hep-ph/0109036}.

\bibitem[Kawabata(1995)]{Kawabata:1995th}
S.~Kawabata, \emph{Comp. Phys. Commun.}, \textbf{88}, 309--326 (1995).

\bibitem[Ilyin et~al.(1996)]{Ilyin:1996gy}
V.~A. Ilyin, D.~N. Kovalenko, and A.~E. Pukhov, \emph{Int. J. Mod. Phys.},
  \textbf{C7}, 761 (1996), \eprint{hep-ph/9612479}.

\bibitem[Mangano et~al.(2003)]{Mangano:2002ea}
M.~L. Mangano, M.~Moretti, F.~Piccinini, R.~Pittau, and A.~D. Polosa,
  \emph{JHEP}, \textbf{07}, 001 (2003), \eprint{hep-ph/0206293}.

\bibitem[Campbell and Ellis(2000)]{Campbell:2000bg}
J.~M. Campbell, and R.~K. Ellis, \emph{Phys. Rev.}, \textbf{D62}, 114012
  (2000), \eprint{hep-ph/0006304}.

\bibitem[Beenakker et~al.(1996)]{Beenakker:1996ed}
W.~Beenakker, R.~Hopker, and M.~Spira (1996), \eprint{hep-ph/9611232}.

\bibitem[Berends et~al.(1989)]{Berends:1989ie}
F.~A. Berends, W.~T. Giele, and H.~Kuijf, \emph{Phys. Lett.}, \textbf{B232},
  266 (1989).

\bibitem[Berends et~al.(1991)]{Berends:1990ax}
F.~A. Berends, H.~Kuijf, B.~Tausk, and W.~T. Giele, \emph{Nucl. Phys.},
  \textbf{B357}, 32--64 (1991).

\bibitem[Giele et~al.(1993)]{Giele:1993dj}
W.~T. Giele, E.~W.~N. Glover, and D.~A. Kosower, \emph{Nucl. Phys.},
  \textbf{B403}, 633--670 (1993), \eprint{hep-ph/9302225}.

\bibitem[Kersevan and Richter-Was(2003)]{Kersevan:2002dd}
B.~P. Kersevan, and E.~Richter-Was, \emph{Comput. Phys. Commun.}, \textbf{149},
  142--194 (2003), \eprint{hep-ph/0201302}.

\bibitem[Nagy and Trocsanyi(1999)]{Nagy:1998bb}
Z.~Nagy, and Z.~Trocsanyi, \emph{Phys. Rev.}, \textbf{D59}, 014020 (1999),
  \eprint{hep-ph/9806317}.

\bibitem[Banfi et~al.(2005)]{Banfi:2004yd}
A.~Banfi, G.~P. Salam, and G.~Zanderighi, \emph{JHEP}, \textbf{03}, 073 (2005),
  \eprint{hep-ph/0407286}.

\bibitem[Amati et~al.(1980)]{Amati:1980ch}
D.~Amati, A.~Bassetto, M.~Ciafaloni, G.~Marchesini, and G.~Veneziano,
  \emph{Nucl. Phys.}, \textbf{B173}, 429 (1980).

\bibitem[Webber(1986)]{Webber:1986mc}
B.~R. Webber, \emph{Ann. Rev. Nucl. Part. Sci.}, \textbf{36}, 253 (1986).

\bibitem[Gustafson and Pettersson(1988)]{Gustafson:1987rq}
G.~Gustafson, and U.~Pettersson, \emph{Nucl. Phys.}, \textbf{B306}, 746 (1988).

\bibitem[L{\"o}nnblad(1992)]{Lonnblad:1992tz}
L.~L{\"o}nnblad, \emph{Comput. Phys. Commun.}, \textbf{71}, 15--31 (1992).

\bibitem[L{\"o}nnblad(1996)]{Lonnblad:1995ex}
L.~L{\"o}nnblad, \emph{Nucl. Phys.}, \textbf{B458}, 215--230 (1996),
  \eprint{hep-ph/9508261}.

\bibitem[Marchesini and Webber(1984)]{Marchesini:1983bm}
G.~Marchesini, and B.~R. Webber, \emph{Nucl. Phys.}, \textbf{B238}, 1 (1984).

\bibitem[Seymour(1995)]{Seymour:1994df}
M.~H. Seymour, \emph{Comp. Phys. Commun.}, \textbf{90}, 95--101 (1995),
  \eprint{hep-ph/9410414}.

\bibitem[Corcella et~al.(2001)]{Corcella:2000bw}
G.~Corcella, et~al., \emph{JHEP}, \textbf{01}, 010 (2001),
  \eprint{hep-ph/0011363}.

\bibitem[Gieseke et~al.(2003)]{Gieseke:2003rz}
S.~Gieseke, P.~Stephens, and B.~Webber, \emph{JHEP}, \textbf{12}, 045 (2003),
  \eprint{hep-ph/0310083}.

\bibitem[Bengtsson and Sj{\"o}strand(1987)]{Bengtsson:1986et}
M.~Bengtsson, and T.~Sj{\"o}strand, \emph{Nucl. Phys.}, \textbf{B289}, 810
  (1987).

\bibitem[Norrbin and Sj{\"o}strand(2001)]{Norrbin:2000uu}
E.~Norrbin, and T.~Sj{\"o}strand, \emph{Nucl. Phys.}, \textbf{B603}, 297--342
  (2001), \eprint{hep-ph/0010012}.

\bibitem[Sj{\"o}strand et~al.(2001)]{Sjostrand:2000wi}
T.~Sj{\"o}strand, et~al., \emph{Comput. Phys. Commun.}, \textbf{135}, 238--259
  (2001), \eprint{hep-ph/0010017}.

\bibitem[Sj{\"o}strand and Skands(2005)]{Sjostrand:2004ef}
T.~Sj{\"o}strand, and P.~Z. Skands, \emph{Eur. Phys. J.}, \textbf{C39},
  129--154 (2005), \eprint{hep-ph/0408302}.

\bibitem[Gleisberg et~al.(2004)]{Gleisberg:2003xi}
T.~Gleisberg, et~al., \emph{JHEP}, \textbf{02}, 056 (2004),
  \eprint{hep-ph/0311263}.

\bibitem[Sj{\"o}strand and Skands(2003)]{Sjostrand:2002ip}
T.~Sj{\"o}strand, and P.~Z. Skands, \emph{Nucl. Phys.}, \textbf{B659}, 243
  (2003), \eprint{hep-ph/0212264}.

\bibitem[Corcella and Moretti(2004)]{Corcella:2004fu}
G.~Corcella, and S.~Moretti (2004), \eprint{hep-ph/0402149}.

\bibitem[Miu and Sj{\"o}strand(1999)]{Miu:1998ju}
G.~Miu, and T.~Sj{\"o}strand, \emph{Phys. Lett.}, \textbf{B449}, 313--320
  (1999), \eprint{hep-ph/9812455}.

\bibitem[Andre and Sj{\"o}strand(1998)]{Andre:1997vh}
J.~Andre, and T.~Sj{\"o}strand, \emph{Phys. Rev.}, \textbf{D57}, 5767--5772
  (1998), \eprint{hep-ph/9708390}.

\bibitem[Mrenna(1999)]{Mrenna:1999mq}
S.~Mrenna (1999), \eprint{hep-ph/9902471}.

\bibitem[Mangano(2004)]{Mangano}
M.~Mangano, {"The so--called MLM prescription for ME/PS matching"} (2004),
  {Talk presented at the Fermilab ME/MC Tuning Workshop, October 4, 2004}.

\bibitem[Catani et~al.(2001)]{Catani:2001cc}
S.~Catani, F.~Krauss, R.~Kuhn, and B.~R. Webber, \emph{JHEP}, \textbf{11}, 063
  (2001), \eprint{hep-ph/0109231}.

\bibitem[Krauss(2002)]{Krauss:2002up}
F.~Krauss, \emph{JHEP}, \textbf{08}, 015 (2002), \eprint{hep-ph/0205283}.

\bibitem[L{\"o}nnblad(2002)]{Lonnblad:2001iq}
L.~L{\"o}nnblad, \emph{JHEP}, \textbf{05}, 046 (2002), \eprint{hep-ph/0112284}.

\bibitem[Mrenna and Richardson(2004)]{Mrenna:2003if}
S.~Mrenna, and P.~Richardson, \emph{JHEP}, \textbf{05}, 040 (2004),
  \eprint{hep-ph/0312274}.

\bibitem[Frixione(2005)]{dis_frixione}
S.~Frixione, \enquote{Monte-{C}arlo generators,} in \emph{these proceedings},
  2005, transparencies available from www.hep.wisc.edu/dis05/.

\bibitem[Lavesson and L{\"o}nnblad(2005)]{Lavesson:2005xu}
N.~Lavesson, and L.~L{\"o}nnblad (2005), \eprint{hep-ph/0503293}.

\bibitem[Mrenna(2004)]{patriot}
S.~Mrenna, Patriot database (2004), see {\\
  \texttt{http://cepa.fnal.gov/personal/mrenna/Matched\_Dataset\_Description.h%
tml}}.

\bibitem[Dobbs(2002)]{Dobbs:2001dq}
M.~Dobbs, \emph{Phys. Rev.}, \textbf{D65}, 094011 (2002),
  \eprint{hep-ph/0111234}.

\bibitem[Potter and Schorner(2001)]{Potter:2001ej}
B.~Potter, and T.~Schorner, \emph{Phys. Lett.}, \textbf{B517}, 86--92 (2001),
  \eprint{hep-ph/0104261}.

\bibitem[Collins(2000)]{Collins:2000qd}
J.~C. Collins, \emph{JHEP}, \textbf{05}, 004 (2000), \eprint{hep-ph/0001040}.

\bibitem[Collins and Hautmann(2001)]{Collins:2000gd}
J.~C. Collins, and F.~Hautmann, \emph{JHEP}, \textbf{03}, 016 (2001),
  \eprint{hep-ph/0009286}.

\bibitem[Collins and Zu(2005)]{Collins:2004vq}
J.~C. Collins, and X.~Zu, \emph{JHEP}, \textbf{03}, 059 (2005),
  \eprint{hep-ph/0411332}.

\bibitem[Frixione and Webber(2002)]{Frixione:2002ik}
S.~Frixione, and B.~R. Webber, \emph{JHEP}, \textbf{06}, 029 (2002),
  \eprint{hep-ph/0204244}.

\bibitem[Frixione et~al.(2003)]{Frixione:2003ei}
S.~Frixione, P.~Nason, and B.~R. Webber, \emph{JHEP}, \textbf{08}, 007 (2003),
  \eprint{hep-ph/0305252}.

\bibitem[Kramer and Soper(2004)]{Kramer:2003jk}
M.~Kramer, and D.~E. Soper, \emph{Phys. Rev.}, \textbf{D69}, 054019 (2004),
  \eprint{hep-ph/0306222}.

\bibitem[Soper(2004)]{Soper:2003ya}
D.~E. Soper, \emph{Phys. Rev.}, \textbf{D69}, 054020 (2004),
  \eprint{hep-ph/0306268}.

\bibitem[Nagy and Soper(2005)]{Nagy:2005aa}
Z.~Nagy, and D.~E. Soper (2005), \eprint{hep-ph/0503053}.

\bibitem[Gieseke et~al.(2004)]{Gieseke:2003hm}
S.~Gieseke, A.~Ribon, M.~H. Seymour, P.~Stephens, and B.~Webber, \emph{JHEP},
  \textbf{02}, 005 (2004), \eprint{hep-ph/0311208}.

\bibitem[Sj{\"o}strand et~al.(2003)]{Sjostrand:2003wg}
T.~Sj{\"o}strand, L.~L{\"o}nnblad, S.~Mrenna, and P.~Skands (2003),
  \eprint{hep-ph/0308153}.

\bibitem[Krauss et~al.(2005)]{Krauss:2005re}
F.~Krauss, A.~Schalicke, and G.~Soff (2005), \eprint{hep-ph/0503087}.

\bibitem[Webber(1984)]{Webber:1983if}
B.~R. Webber, \emph{Nucl. Phys.}, \textbf{B238}, 492 (1984).

\bibitem[Gieseke(2004)]{Gieseke:2004af}
S.~Gieseke (2004), \eprint{hep-ph/0408034}.

\bibitem[Gieseke(2005)]{Gieseke:2004tc}
S.~Gieseke, \emph{JHEP}, \textbf{01}, 058 (2005), \eprint{hep-ph/0412342}.

\bibitem[Gustafson(1986)]{Gustafson:1986db}
G.~Gustafson, \emph{Phys. Lett.}, \textbf{B175}, 453 (1986).

\bibitem[Rudolph(2003)]{Rudolph}
G.~Rudolph (2003), {ALEPH}, unpublished.

\bibitem[Huston et~al.(2004)]{Huston:2004yp}
J.~Huston, I.~Puljak, T.~Sj{\"o}strand, and E.~Thom{\'e}, \enquote{Resummation
  and shower studies,} in \emph{The QCD/SM working group: Summary report, 3rd
  Les Houches Workshop: Physics at TeV Colliders, Les Houches, France, 26 May -
  6 Jun 2003, M.~Dobbs {\it et al.}, hep-ph/0403100}, 2004,
  \eprint{hep-ph/0401145}.

\bibitem[Plehn et~al.(2005)]{plehnprep}
T.~Plehn, D.~Rainwater, and P.~Skands (2005), in preparation.

\bibitem[Field(2002--2003)]{Field}
R.~Field, presentations at the {`Matrix Element and Monte Carlo Tuning
  Workshop}, {Fermilab, 4 October 2002 and 29--30 April 2003} (2002--2003),
  talks available from webpage \texttt{http://cepa.fnal.gov/CPD/MCTuning/}, and
  further recent talks available from
  \texttt{http://www.phys.ufl.edu/}$\sim$\texttt{rfield/cdf/}.

\bibitem[Sj{\"o}strand and van Zijl(1987)]{Sjostrand:1987su}
T.~Sj{\"o}strand, and M.~van Zijl, \emph{Phys. Rev.}, \textbf{D36}, 2019
  (1987).

\bibitem[Butterworth et~al.(1996)]{Butterworth:1996zw}
J.~M. Butterworth, J.~R. Forshaw, and M.~H. Seymour, \emph{Z. Phys.},
  \textbf{C72}, 637--646 (1996), \eprint{hep-ph/9601371}.

\bibitem[Sj{\"o}strand and Skands(2004)]{Sjostrand:2004pf}
T.~Sj{\"o}strand, and P.~Z. Skands, \emph{JHEP}, \textbf{03}, 053 (2004),
  \eprint{hep-ph/0402078}.

\bibitem[Schumann(2004)]{uesherpa}
S.~Schumann, {SHERPA}: an event generator for the {LHC} (2004), talk given at
  {HERA/LHC Workshop}, {CERN}, {October 2004}.

\bibitem[Boos et~al.(2001)]{Boos:2001cv}
E.~Boos, et~al. (2001), \eprint{hep-ph/0109068}.

\bibitem[Skands et~al.(2004)]{Skands:2003cj}
P.~Skands, et~al., \emph{JHEP}, \textbf{07}, 036 (2004),
  \eprint{hep-ph/0311123}.

\end{thebibliography}

\IfFileExists{\jobname.bbl}{}
 {\typeout{}
  \typeout{******************************************}
  \typeout{** Please run "bibtex \jobname" to obtain}
  \typeout{** the bibliography and then re-run LaTeX}
  \typeout{** twice to fix the references!}
  \typeout{******************************************}
  \typeout{}
 }

\end{document}